\documentclass[prl,superscriptaddress,twocolumn,showkeys,preprintnumbers,floats,floatfix]{revtex4}

\usepackage{graphicx}
\usepackage{amsmath}

\begin{document}

\title{Spin filling of valley-orbit states in a silicon quantum dot}

\author{W. H. Lim}
\email[Electronic mail: ]{wee.lim@unsw.edu.au}
\author{C. H. Yang}
\author{F. A. Zwanenburg}

\author{A. S. Dzurak}
\affiliation{Centre for Quantum Computation and Communication Technology, School of Electrical Engineering \& Telecommunications, The University of New South Wales, Sydney 2052, Australia}

\date{\today}

\begin{abstract}
We report the demonstration of a low-disorder silicon
metal-oxide-semiconductor (Si MOS) quantum dot containing a tunable
number of electrons from zero to $N=27$. The observed evolution of
addition energies with parallel magnetic field reveals the spin
filling of electrons into valley-orbit states. We find a splitting
of 0.10~meV between the ground and first excited states, consistent
with theory and placing a lower bound on the valley splitting. Our
results provide optimism for the realization in the near future of
spin qubits based on silicon quantum dots.
\end{abstract}

\pacs{71.55.-i, 73.20.-r, 76.30.-v, 84.40.Az, 85.40.Ry}

\keywords{quantum dot, silicon, spin filling, valley splitting}

\maketitle

\textbf{1. Introduction}

Semiconductor quantum dots~\cite{Kouwenhoven1997} are islands to
which electrons can be added one by one by means of an electric
field. Like real atoms they have discrete quantum levels and can
exhibit phenomena such as shell filling~\cite{Tarucha1996}, where
orbital levels are filled by spin-paired electrons to produce a
spin-zero many-electron state. Quantum dots also provide a promising platform for spin qubits, which can have long coherence times due to the weak coupling of spins to local fluctuations in charge. For a quantum dot to be useful as a
spin qubit it is essential to understand the details of its
excitation spectrum and its spin-filling structure. One powerful
method to probe the spin filling is via magnetospectroscopy. This
has been applied to both vertical~\cite{Tarucha1996} and lateral
GaAs/AlGaAs quantum dots~\cite{Ciorga2000}, showing ground-state
spin filling in agreement with Hund's rule.

Silicon devices are attractive for spin-based quantum
computing~\cite{Loss1998,Kane1998} and spintronics~\cite{Zutic2004}
because of their scalability and long spin coherence
times~\cite{Tyryshkin2003}. Silicon quantum dots, in particular,
have potential as electron-spin qubits, but to date it has not been
possible to create devices with the low disorder present in their
GaAs counterparts~\cite{Tarucha1996,Ciorga2000}. This is primarily
due to disorder at the Si/SiO$_2$ interface, which has made it
difficult to achieve single-electron
occupancy~\cite{Lim2009,Xiao2010}.

In addition, the conduction band structure in silicon is complex and
only a few experiments have been carried out to examine the spin
states in either Si MOS or Si/SiGe quantum
dots~\cite{Xiao2010,Rokhinson2001,Hu2009,borselli2010,Simmons2010}.
The valley degree of freedom makes the measurement and
interpretation of spin states in all silicon-based dots
non-trivial~\cite{Hada2003,Culcer2010}, while for Si MOS dots the
substantial amount of disorder usually present at the Si/SiO$_2$
interface impedes the ability to make smooth potential wells.

In this work we present the investigation of a Si MOS quantum dot
with lower disorder than any studied to date, in which it is
possible to analyse the electron occupancy in a manner previously
inaccessible. We deduce the spin filling of the first 12 electrons
in the dot from ground-state magnetospectroscopy measurements. The
formation of a two-electron ($N=2$) spin-singlet state at low
magnetic fields confirms that there is no valley degeneracy present,
while the magnetic field dependence of the higher-order Coulomb
peaks allows us to deduce the level structure for the first four
electrons.

In the following section we present the architecture of the quantum dot and the charge stability diagram in the few-electron regime. We then, on section 3, study the valley-orbit states in this quantum dot and extract a valley-orbit splitting of 0.10 meV. In section 4 we investigate the spin filling of the first 4 electrons in this quantum dot in detail. We then present the spin filling of the 5$^{\textrm{th}}$ to 12$^{\textrm{th}}$ electrons, discussing some anomalies observed, before concluding in section 6. \\

\begin{figure}[tb!]
\includegraphics[width=9.0cm]{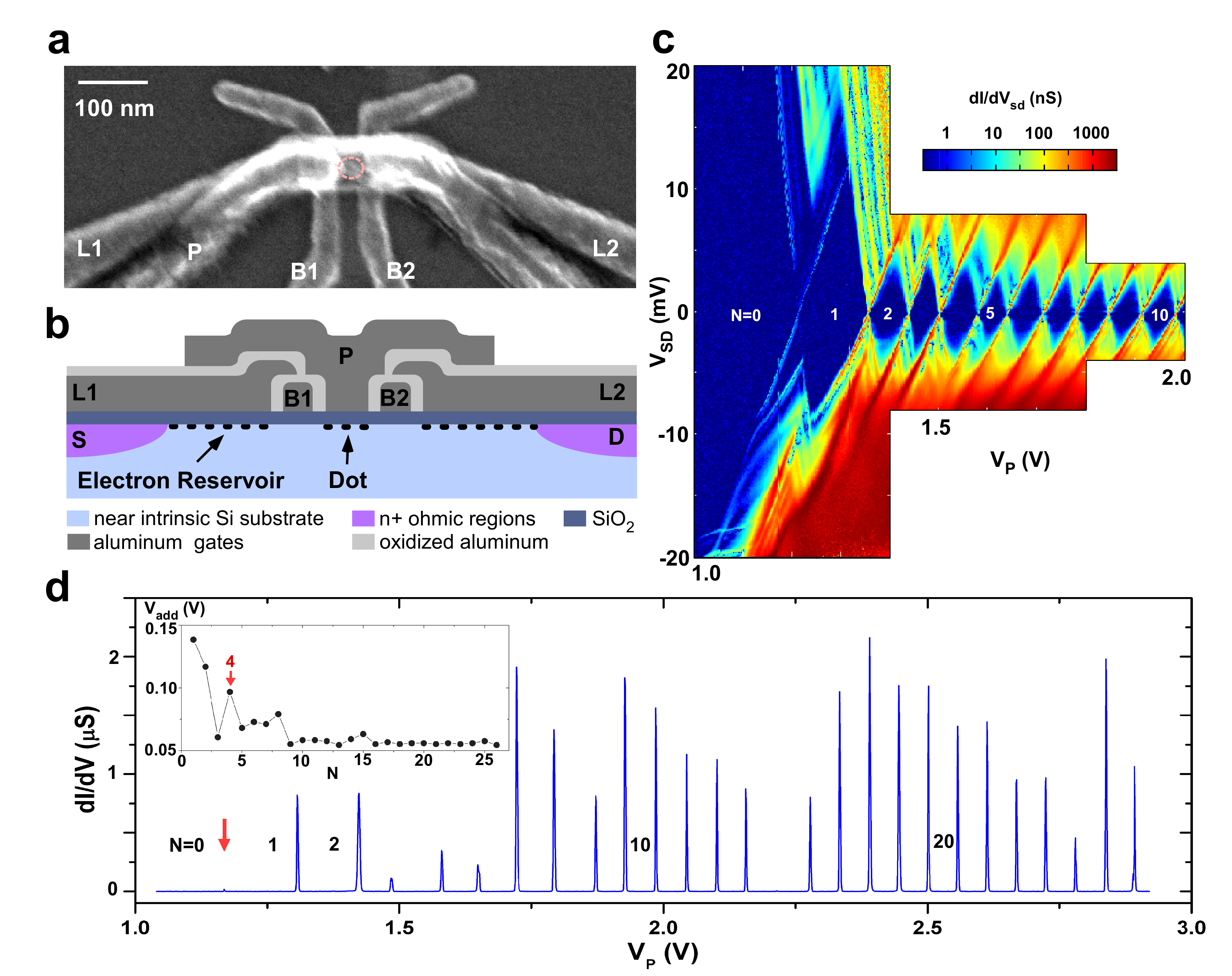}
\caption{\textbf{a}, Scanning electron microscope image and
\textbf{b}, Schematic cross-section of a Si MOS quantum dot.
\textbf{c}, Stability diagram of the device in the few-electron
regime. By decreasing the plunger gate voltage $V_{\textrm{P}}$,
electrons are depleted one-by-one from the dot. The first diamond
opens up completely indicating that the first electron has tunneled
off the dot. \textbf{d}, Coulomb oscillations as a function of
plunger gate voltage $V_{\textrm{P}}$ for the first 27 electrons in
the dot. $V_{\textrm{P}}$ is compensated by $V_\textrm{B2}$ to
suppress the non-monotonic background conductance. Inset: Addition
voltage $V_{\textrm{add}}$ versus electron number $N$ calculated as
the difference between two consecutive Coulomb peaks in plunger gate
voltage.} \label{fig1}
\end{figure}

\textbf{2. Low-disorder silicon MOS quantum dot}

The triple-layer gate stack in our structure (Figure 1a and 1b)
provides excellent flexibility for tuning the barrier transparency
and the energy levels of the dot independently, see Supplementary Information for fabrication processes. The lowest layer defines the barrier gates (B1 and B2). They are used to define the dot spatially and control the tunnel coupling. The second layer of gates defines the soure-drain leads (L1 and L2). The lead gates induce the electron accumulation layers that act as source-drain reservoirs. The plunger gate (P) extends over the barrier gates, lead gates and the dot island, and is used to control the electron occupancy of the dot. Figure 1c is a plot
of the differential conductance $dI/dV_\textrm{SD}$ of the device
versus plunger gate voltage $V_\textrm{P}$ and source-drain voltage
$V_\textrm{SD}$, showing the familiar ``Coulomb diamond" charge
stability map. Before the first charge transition the diamond edges
open entirely to a source-drain voltage $|V_\textrm{SD}|>$ 20 mV,
because the quantum dot has been fully depleted of electrons. We
have previously reported a device with similar gate architecture but
an accidental parallel quantum dot created distortion of the charge
stability map in the few-electron regime, complicating the
interpretation of the dot's level structure~\cite{Lim2009}. Here,
clear and sharp Coulomb peaks mark the first 27 electrons entering
the dot, see Fig. 1d, while the charge stability map of Fig. 1c
shows no distortions from disorder potentials.

As with quantum dots in GaAs/AlGaAs~\cite{Tarucha1996}, shell
filling has very recently been observed in Si/SiGe quantum dots,
with a filled shell structure observed for $N=4$
electrons~\cite{borselli2010}. The addition spectrum of our Si MOS
quantum dot (inset of Figure 1d) also shows a noticeable peak at
$N=4$. A filled shell at $N=4$ would be consistent with the filling
of a first orbital state in a two-valley system, however, an
accurate description of orbital and valley levels in silicon quantum
dots is somewhat more complex, as described below.\\


\textbf{3. Valley-orbit splitting}

In recent years, valley physics in silicon has been studied
extensively both
theoretically~\cite{Boykin1994,Friesen22010,Saraiva22010,Culcer2010,Saraiva2009,Friesen2010,Saraiva2010,Culcer22010}
and
experimentally~\cite{kohler1979,Takashina2006,Goswami2007,Fuechsle2010}.
In bulk silicon, there are six degenerate conduction band minima
(valleys) in the Brillouin zone, as depicted in Figure 2a.
Confinement of electrons in the $z$-direction at the Si/SiO$_2$
interface lifts the six-fold valley degeneracy: four
$\Delta$-valleys with a heavy effective mass parallel to the
interface have an energy several tens of meV higher than the two
$\Gamma$-valleys~\cite{Ando1982}. The sharp and flat interface
produces a potential step in the $z$-direction and lifts the
degeneracy of the $\Gamma$-valleys in two levels separated by the
valley splitting $E_V$. Theoretical predictions for the valley
splitting are generally on the order of 0.1--0.3
meV~\cite{Culcer2010,Saraiva2010}. Experimental values in Si
inversion layers mostly vary from 0.3--1.2 meV~\cite{kohler1979}. A
very large valley splitting of 23~meV in a similar structure has
also been measured~\cite{Takashina2006} and is explained
in~\cite{Saraiva22010,Saraiva2010}. Recently, resonant tunneling
features spaced by $\sim$0.1~meV in a single-crystal silicon quantum
dot were attributed to valley excited states~\cite{Fuechsle2010},
while measurements on Si/SiGe quantum dots revealed valley
splittings in the range of 0.12--0.27~meV~\cite{borselli2010}.

Valleys and orbits can also hybridise~\cite{Friesen2010}, making it
inappropriate to define distinct orbital and valley quantum numbers.
Depending on the degree of mixing, the valley-orbit levels behave
mostly like valleys or like orbits. Instead of referring to a pure
valley splitting we therefore adopt the term valley-orbit splitting,
$\Delta E_{\textrm{VO}}=E_{\textrm{VO2}}-E_{\textrm{VO1}}$ for the
difference in energy between the first two single-particle levels,
$E_{\textrm{VO1}}$ and $E_{\textrm{VO2}}$. This is sometimes
referred to as the ground-state gap~\cite{Friesen2010}.

Full electrostatic control of the electron number allows us to
investigate the spin filling by measuring the magnetic field
dependence of the electrochemical potential $\mu_N$, which is by
definition the energy required for adding the $N^{th}$ electron to
the dot. The slope of $\mu_N(B)$ is given by~\cite{Hada2003}
\begin{equation}
\frac{\partial \mu_N}{\partial B} = -g\mu_B \Delta
S_{\textrm{tot}}(N),
\end{equation}
where $g$ is the g-factor, the Bohr magneton $\mu_\textrm{B} = 58$
$\mu$eV/T and $\Delta S_{\textrm{tot}}(N)$ is the change in total
spin of the dot when the $N^{\textrm{th}}$ electron is added. The
electrochemical potential has a slope of $+g\mu_\textrm{B}/2$ when a
spin-up electron is added, whereas addition of a spin-down electron
results in a slope of $-g\mu_\textrm{B}/2$. The rate at which
$\mu_N$ changes with magnetic field thus reveals the sign of the
added spin. For the experiments in this work we apply the magnetic
field $B~parallel$ to the Si/SiO$_2$ interface.

\begin{figure}[t]
\includegraphics[width=9.5cm]{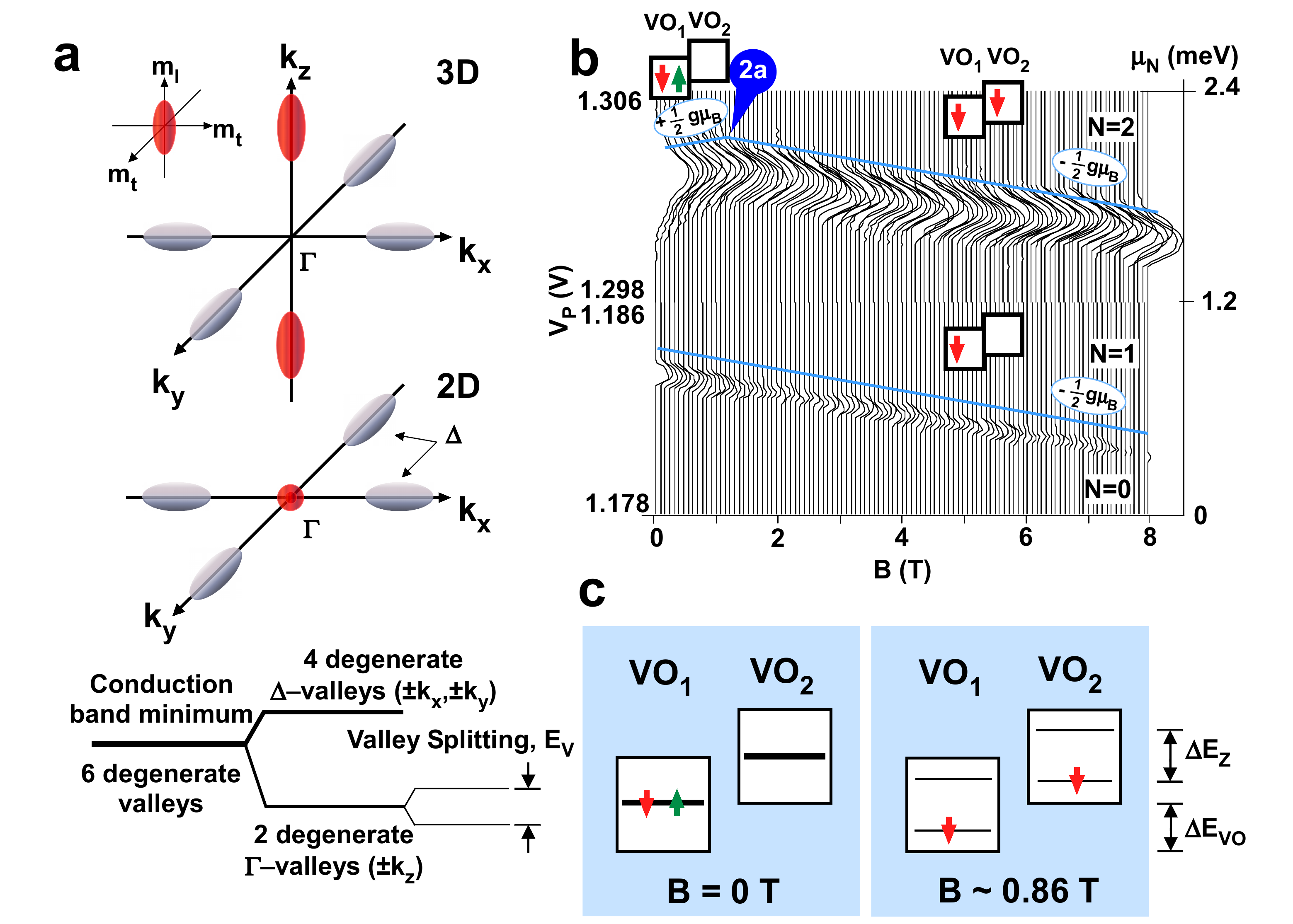}
\caption{\textbf{a}, Conduction band minima (valleys) in bulk silicon, showing six ellipsoids of equal energy in the Brillouin
zone. Each ellipsoid has two light traverse mass (m$_t$) and a heavy
longitudinal mass (m$_l$). Under the $z$-direction confinement at
the Si/SiO$_2$ interface, the six-fold degenerate valleys split into
two $\Gamma$-valleys (lower in energy) and four $\Delta$-valleys
(higher in energy). The sharp interface potentials split the
$\Gamma$-valleys by an amount $E_V$. \textbf{b}, Magnetospectroscopy
of the first two electrons entering the quantum dot. The circle 2a
marks a kink in the second Coulomb peak at $\sim$0.86 T. The arrows
in the boxes (VO$_1$ for valley-orbit 1 and VO$_2$ for valley-orbit
2) represent the spin filling of electrons in the quantum dot.
Coulomb peak positions in gate voltage are converted to energies
using the lever arm $\alpha_\textrm{P}$ extracted from the
corresponding Coulomb diamonds. \textbf{c}, A model showing that the
valley-orbit splitting can be estimated from the magnetic field at which
$\Delta E_{\textrm{VO}}=\Delta E_\textrm{Z}$, i.e. when the spin-up state of VO$_1$
is at the same energy as the spin-down state of VO$_2$. For $B<0.86$
T, the first two electrons fill with opposite spins in the same
valley-orbit level (left panel). As we increase the magnetic field,
the Zeeman energy exceeds the valley-orbit splitting and the second
electron occupies a spin-down state in valley-orbit 2. The sign
change appears as a kink and occurs when the valley-orbit splitting
is equal to the Zeeman energy (0.10~meV).} \label{fig2}
\end{figure}

The conductance at the first two charge transitions is plotted as a
function of the electrochemical potential energy and the magnetic
field in Figure 2b. Here, the Coulomb peak positions in gate voltage
are converted to electrochemical potential $\mu_N$ using the lever
arm $\alpha_\textrm{P}$ extracted from the corresponding Coulomb
diamonds. The blue lines above the Coulomb peaks are guides for the
eye with slopes of $\pm g\mu_\textrm{B}/2$, as predicted by equation
(1) using $g$ = 2 for bulk silicon. Since the first Coulomb peak
moves down in energy with increasing magnetic field the peak
corresponds to a spin-down electron entering the quantum dot, as
expected for the $N=1$ ground state. For $B\geq1$~T the second
Coulomb peak also falls in energy with increasing $B$ at a rate
close to $-g\mu_\textrm{B}/2$, however, for low magnetic fields the
peak noticeably increases in energy with $B$, leading to a ``kink"
(marked 2a) at $B\sim0.86$~T. This kink (2a) is confirmed by several
repeated measurements over positive and negative magnetic field, see
Supplementary Information (Fig. S2). These results imply that at low magnetic
field (before the kink), the second electron fills the quantum dot
with its spin up. As we increase the magnetic field (after the
kink), the sign of the second electron spin changes from up to down
at $B\sim0.86$~T. We note that in previous measurements on a similar
quantum dot device, disorder and instability made it difficult to accurately probe this kink feature~\cite{Lim2009}.

We explain the sign change observed here with a simple model where
the two lowest valley-orbit levels are separated by the valley-orbit
splitting $\Delta E_{\textrm{VO}}$, see Figure 2c. At zero magnetic
field, the first two electrons fill with opposite spins in
valley-orbit level 1. When a magnetic field is applied, the
spin-down and spin-up states are split by the Zeeman energy $E_Z$.
Above 0.86 T the spin-up state of valley-orbit level 1 (VO$_1$) is
higher in energy than the spin-down state of valley-orbit level 2
(VO$_2$) and it becomes energetically favoured for the second
electron to occupy the latter, i.e. VO$_2$. At the kink the
valley-orbit splitting equals the Zeeman energy, which is 0.10 meV
at 0.86 T. With the interfacial electric field of
$\sim$2$\times10^7$~V/m extracted from Technology
Computer-Aided-Design modeling for our device structure, the
valley-orbit splitting agrees well with modeling results
(0.08--0.11~meV) based on the effective-mass
approximation~\cite{Saraiva2010,Culcer22010}. We note that if no
valley-orbit mixing were present, then $\Delta
E_{\textrm{VO}}=0.10$~meV would place a lower bound on the valley
splitting for this structure.

For $B > 0.86$ T the first two electrons fill two different levels
split by $\Delta E_{\textrm{VO}}$ = 0.10 meV. We note that the
presence of a doubly degenerate ground-state level would demand the
two electrons to exhibit parallel spin filling starting from 0 T,
since the two electrons would then occupy two different valley
states in order to minimise the exchange energy~\cite{Hada2003}.
\emph{A valley-degenerate state is therefore ruled out by the
results in Figure 2b}.

To assess the degree of valley-orbit mixing we compare the expected
values for the orbital level spacing and the valley splitting. As
stated above, theoretical calculations of the latter predict
0.1--0.3 meV. An estimate of the orbital level spacing in a quantum
dot is given by $2\pi\hbar^2/g_vg_sm^*A$ \cite{Kouwenhoven1997},
where $g_v$ ($g_s$) is the valley (spin) degeneracy, $m^*$ the
electron effective mass and $A$ the dot area. For non-degenerate
valleys, $g_v=1$ and $g_s=2$. Using the effective mass of $0.19m_0$,
and the lithographic dot area of $\sim30\times60$ nm$^2$ we obtain
an expected orbital level spacing of 0.7 meV. This value is
considerably larger than the lower bound on the valley splitting,
suggesting that the first two levels may be valley-like, however, to
maintain generality we will continue to refer to the levels as
valley-orbit states.\\

\textbf{4. Spin filling of the first 4 electrons}

We now turn to the spin filling for $N\geq2$ electrons. Figure 3a
shows the differential conductance as a function of plunger gate
voltage and barrier gate voltage $V_\textrm{B2}$. The highly regular
pattern of parallel Coulomb peak lines again demonstrates the low
disorder in this device. In order to determine the spin filling for
higher electron numbers we investigate the \emph{difference} between
successive electrochemical potentials as a function of magnetic
field. The resulting addition energies $E_{\textrm{add}}(N) =
\mu_N-\mu_{N-1}$ have slopes which depend on the spin filling of two
consecutive electrons, according to~\cite{hanson2007}
\begin{align}
\frac{\partial E_{\textrm{add}}(N)}{\partial B} &= 0 \hspace{+1.76cm} \textrm{for} \downarrow,\downarrow \textrm{or} \uparrow,\uparrow \\
\nonumber &=-g\mu_B  \hspace{+1.0cm} \textrm{for} \uparrow,\downarrow\\
\nonumber &=+g\mu_B  \hspace{+1.0cm} \textrm{for}
\downarrow,\uparrow
\end{align}
where the first (second) arrow depicts the spin of the
$(N-1)^{\textrm{th}}$~($N^{\textrm{th}}$) electron respectively.

\begin{figure}[t] 
\includegraphics[width=9.0cm]{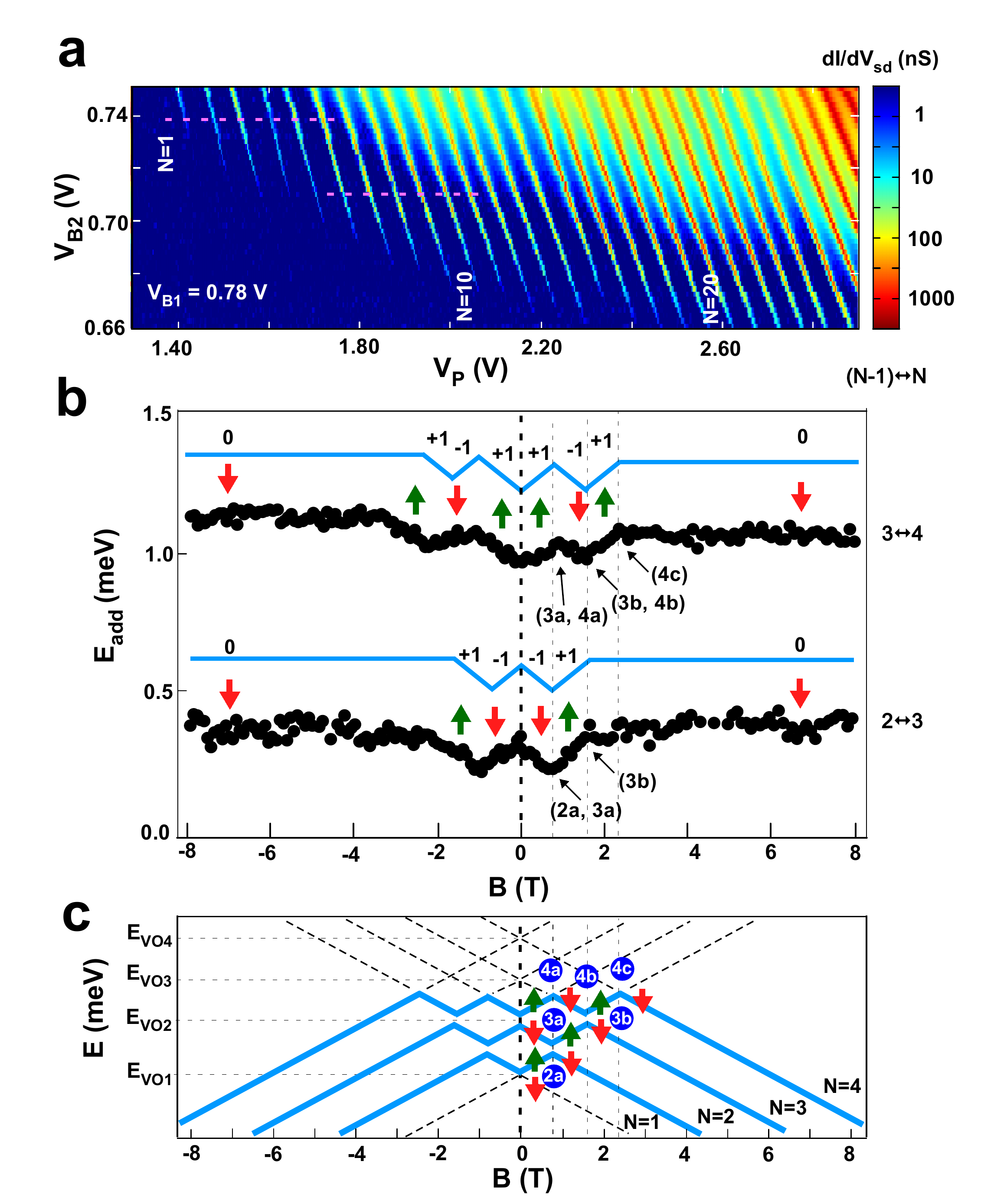} 
\caption{\textbf{a}, Differential conductance d$I$/d$V_\textrm{sd}$
as a function of barrier gate voltage $V_{\textrm{B2}}$ and plunger
gate voltage $V_{\textrm{P}}$ at $B=0$~T. The regular parallel
Coulomb peaks are a signature of low disorder. \textbf{b}, Addition energies of the
3$^{\textrm{rd}}$ and 4$^{\textrm{th}}$ electrons versus magnetic
field. Kinks are reproducible and approximately symmetric over positive and
negative magnetic fields. \textbf{c}, A simple model showing the
evolution of single-particle energy levels $E_{\textrm{VO}i}$ of
valley-orbit $i$ assuming only the Zeeman shift. Each level splits
into two levels $E_{\textrm{VO}i} \pm\frac{1}{2}g\mu_\textrm{B}B$ at
non-zero magnetic fields. The level crossings fit the kinks observed
in the first four Coulomb peaks shown in Fig. 2b and Fig. 3b.}
\label{fig3}
\end{figure}

Figure 3b plots the measured addition energies, $E_{\textrm{add}}(N)
= \mu_N-\mu_{N-1}$, for $N=2$ to $N=4$ electrons for magnetic fields
$B$ in the range of $-8$~T$<B<8$~T. We see that the data in Fig. 3b
tend to follow $\partial E_{\textrm{add}}(N)/{\partial B}= 0,
\pm g\mu_B$, as expected from Equation (2). Furthermore, the
$E_{\textrm{add}}(N)$ data is relatively symmetric about $B=0$, indicating that the trends are real and not
measurement artefacts. As a guide to the eye, we also show lines
with slopes of \emph{exactly} 0, $\pm g\mu_B$ (blue lines in Fig.
3b) that we interpret the $E_{\textrm{add}}(N)$ to be following.
While in regions the match is not exact, we propose that
these trend lines are the best qualitative fit to the data. We are thus able to infer spin states for each of the first 4 electrons at all values of magnetic field $|B|<8$~T. These spin
states are labelled with red (green) arrows, representing spin down
(up), in Fig. 3b.


We now focus on the spin states of the these four electrons, $N=1$
to $N=4$. At low magnetic fields ($<0.8$~T), the electrons populate
the quantum dot ground states with alternating spin directions:
$\downarrow,\uparrow,\downarrow,\uparrow$. Conversely, at high
magnetic fields ($>4$ T) a configuration with four spin-down
electrons has least energy:
$\downarrow,\downarrow,\downarrow,\downarrow$. Recently, parallel
spin filling in a Si quantum dot was explained as a result of a
large exchange energy and an unusually large valley splitting of
0.77 meV \cite{Xiao2010}. When the level spacing is smaller than the
exchange energy, it is energetically favoured for two electrons to
occupy two consecutive levels with the same spin sign. This is not
the case for the device measured here: the anti-parallel spin
filling of the first two electrons below 0.86 T is only possible in
case of a small exchange energy (less than $\Delta
E_{\textrm{VO}}$). This is an unexpected result for a dot of this
size where the exchange energy is predicted to be larger than
the orbital level spacing \cite{Hada2003}. Possibly the Coulomb
interaction in the dot is strongly screened by the plunger gate.
This is not unlikely since the distance from gate to dot (10 nm) is
smaller than the dimensions of the dot itself (30$-$60 nm).

In Figure 3c, we illustrate the magnetic-field evolution of four
non-degenerate valley-orbit levels by means of an elementary model.
Each level splits into spin-up and spin-down levels in finite
magnetic field. We assume that the exchange interaction is small in
comparison to the level spacing. The level crossings that follow
from our model fit the kinks observed in the first four Coulomb
peaks. The observed kink positions yield three valley-orbit levels
which are 0.10, 0.23 and 0.29 meV above the lowest ground state
level. The extracted level spacings for the first four valley-orbit
states are then: $E_{\textrm{VO2}}-E_{\textrm{VO1}}=0.10$~meV;
$E_{\textrm{VO3}}-E_{\textrm{VO2}}=0.13$~meV; and
$E_{\textrm{VO4}}-E_{\textrm{VO3}}=0.06$~meV.\\

\begin{figure}[t]
\includegraphics[width=8.0cm]{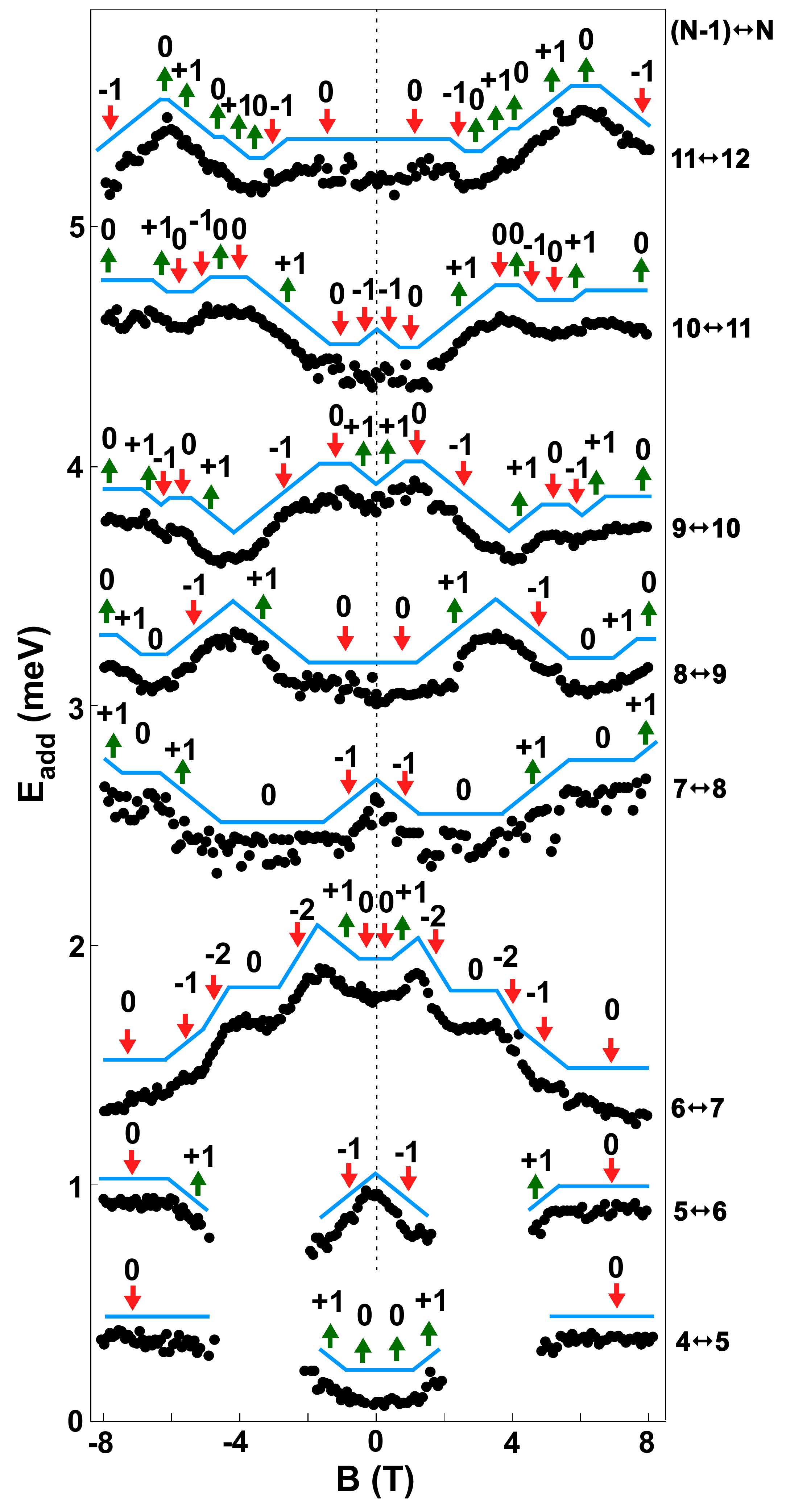}
\caption{Addition energies of the 5$^{\textrm{th}}$ to
12$^{\textrm{th}}$ electron versus magnetic field, offset for
clarity.  Black numbers -2, -1, 0, +1 and +2 correspond to the
slopes of the addition energy in multiples of $g\mu_B$ and reveal
the sign of the added spin (indicated as red and green arrows)
according to equation (2). Coulomb peak spacings in gate voltage are
converted to energies using the lever arm $\alpha_\textrm{P}$, which
vary from 0.11 to 0.068 eV/V with increasing electron number. The
measurement was taken along the dashed line marked in Fig 3(a) at
$V_{\textrm{B2}}$=0.738~V for the 5$^{\textrm{th}}$ to
7$^{\textrm{th}}$ electron. The addition energies of the
8$^{\textrm{th}}$ to 12$^{\textrm{th}}$ electron were taken at
$V_{\textrm{B2}}$=0.710~V.}
\label{fig4}
\end{figure}

\textbf{5. Spin filling of electrons 5--12}

Finally in Figure 4, we plot the addition energies
$E_{\textrm{add}}(N)$ as a function of $B$ for electrons $N=5$ to
12. Once again, we predominantly observe slopes of $\partial
E_{\textrm{add}}(N)/{\partial B}= 0, \pm g\mu_B$, as expected from
Equation (2). Occasionally, e.g. at $N=6\leftrightarrow7$, a segment
has a slope of $\pm 2g\mu_B$, because the total spin on the dot
changes by more than $\frac{1}{2}$. This can occur due to many-body
interactions on the dot and lead to spin
blockade~\cite{Weinmann1995}. The latter phenomenon could also
explain the suppression of current in the fifth charge transition at
$B=2$$-$5~T~\cite{Rokhinson2001,Hu2009,Hada2003}.

Also, the picture of alternating spin filling below 0.8~T no longer holds
for $N>4$. Unexpectedly, the fifth electron is spin up at low
magnetic field, while the lowest-energy configuration predicts a
spin-down state. This anomaly could be explained by an extra
electron in a dot nearby, which alters the spin configuration of the
main dot. Such a small dot can be created at high plunger gate
voltages, where the potential well differs from a perfect parabola. As more electrons are added to the main dot, the wavefunctions extend further and would have more opportunity to spin-couple to the unintentional dot nearby, thus affecting the spin filling of electrons. \\

\textbf{6. Conclusion}

The results here show that silicon MOS quantum dots can be
fabricated with the low levels of disorder necessary to form
well-defined electron spin qubits in a host material that can be
made almost free of nuclear spins. The excellent charge stability
allows the spin states of the dot to be mapped up to $N=12$
electrons and a valley-orbit splitting of 0.10~meV to be extracted.
A recent theoretical study~\cite{Culcer2010} has shown that a valley
splitting of 0.1~meV is sufficient for the operation of a silicon
double quantum dot as a singlet-triplet qubit, in analogy with
recent experiments in GaAs~\cite{Petta2005}. Given that the
valley-orbit splitting is strongly dependent on the interfacial
electric field, it should be possible to further increase the
splitting via appropriate device engineering. Our results therefore
provide real promise for the realization of low-decoherence spin
qubits based upon silicon MOS technology.

We thank M. Eto, R. Okuyama and M. Friesen for valuable discussions
and comments on the manuscript. We thank D.~Barber and
R.~P.~Starrett for technical support. This work was supported by the
Australian Research Council, the Australian Government, and by the
U. S. National Security Agency (NSA) and U. S. Army Research Office
(ARO) (under Contract No. W911NF-08-1-0527).

\end{document}